\documentclass[a4,oneside,english,reqno,12pt]{amsart}
\usepackage{extsizes}
\usepackage{geometry}
\usepackage{amssymb,amsmath,amsthm,amsfonts,mathtools,esint,dsfont,bbm,yhmath}
\usepackage{babel}
\usepackage{appendix}
\usepackage{enumitem}
\usepackage[pdfencoding=auto,psdextra,hidelinks]{hyperref}
\usepackage{bookmark}
\usepackage{cleveref}
\usepackage{crossreftools}
\crefname{subsection}{subsection}{subsections}
\newcommand{\crefdefpart}[2]{%
  \hyperref[#2]{\namecref{#1}~\labelcref*{#1}~\ref*{#2}}%
}
\usepackage{orcidlink}

\newcommand{\wto}{\rightharpoonup}

\numberwithin{equation}{section}

\theoremstyle{plain}
\newtheorem{theorem}{Theorem}[section]

\makeatother

\theoremstyle{remark}
\newtheorem{remark}[theorem]{Remark}

\theoremstyle{plain}
\makeatletter
\newcommand{\neutralize}[1]{\expandafter\let\csname c@#1\endcsname\count@}

\makeatletter
\@namedef{subjclassname@2020}{\textup{2020} Mathematics Subject Classification}
\makeatother

\allowdisplaybreaks
\begin{document}

\title[Homogeneous Bose gases]{Local density approximation and other limit regimes for a homogeneous Bose gas with repulsive three-body interactions in low-dimensional space}

\author[T.A.T. Doan]{Thi Anh Thu Doan}
\address[T.A.T. Doan]{Faculty of Mathematics and Computer Science, University of Science, Ho Chi Minh City, Vietnam; and Vietnam National University, Ho Chi Minh City, Vietnam; and Faculty of Mathematics and Statistics, Ton Duc Thang University, Ho Chi Minh City, Vietnam.}
\email{\href{mailto:doanthianhthu@tdtu.edu.vn}{doanthianhthu@tdtu.edu.vn}}

\author[D.-T. Nguyen]{Dinh-Thi Nguyen\,\orcidlink{0000-0003-4487-5557}}
\address[D.-T. Nguyen]{Faculty of Mathematics and Computer Science, University of Science, Ho Chi Minh City, Vietnam; and Vietnam National University, Ho Chi Minh City, Vietnam.}
\email{\href{mailto:ndthi@hcmus.edu.vn}{ndthi@hcmus.edu.vn}}

\subjclass[2020]{35J10, 35Q55, 81V70, 82D05}
\keywords{Bose--Einstein condensates, Gagliardo--Nirenberg inequality, nonlinear Schr{\"o}\-dinger equation, three-body interaction, two-body interaction}

\begin{abstract}
In the whole space of low spatial dimensions, namely $d \leq 2$, we study the minimizers of an energy functional with an attractive cubic nonlinearity and a repulsive quintic nonlinearity, which describes a quantum Bose gas with a two-body attraction and a three-body repulsion. We prove the existence of minimizers at fixed effective statistics parameter. In the limit of a large effective statistics parameter of the mass-critical nonlinearity (with respect to the kinetic energy), we derive an effective Thomas--Fermi-like model for the homogeneous Bose gas. We also consider other limit regimes depending on the mass-critical nonlinearity.
\end{abstract}

\maketitle
\tableofcontents

\section{Introduction}

This article is focused on analyzing the ground state properties of a homogeneous Bose gas through a nonlinear Schr\"odinger energy functional. This functional has terms representing both an attractive two-body interaction and a repulsive three-body interaction, often modeled by cubic and quintic nonlinearities, respectively. This is particularly relevant to low-dimensional Bose gases, such as in two-dimensional quantum wells or quasi-one-dimensional systems (e.g., Bose gases in tight traps or optical lattices \cite{ParWidMurManFolCirGorShlTheHanBlo-04,KinWenWei-04,BloDalWil-08,PetHolShl-00}). The competing interactions, represented by the attractive two-body and repulsive three-body terms, can stabilize the system and prevent collapse due to attraction alone, leading to the possibility of a stable Bose--Einstein condensate (BEC) with well-defined and nontrivial ground-state properties, such as modified density profiles or novel excitation spectra.

The rigorous derivation of effective nonlinear Schrödinger (Gross--Pitaevskii) functionals from many-body bosonic Hamiltonians is by now classical in the case of two-body interactions (see, e.g., \cite{LieSeiYng-01,CheHol-13,CheHol-17,LewNamRou-17,NamRouSei-16}; see also \cite{LieSeiSolYng-05} for a general discussion). The physical phenomenon of mass concentration in this setting has been observed and rigorously established in several works (see, e.g., \cite{GuoSei-14,LewNamRou-17-proc,NguRou-22,DinNguRou-24}). In recent years, Bose--Einstein condensates (BECs) with three-body interactions have attracted significant attention in both physics and mathematics. Using advanced techniques in mathematical quantum mechanics, a number of works \cite{ChePav-11,Chen-12,Yuan-15,Xie-15,CheHol-19,NamSal-20,LiYao-21,NamRicTri-22a,NamRicTri-22b,NamRicTri-23} have rigorously derived effective nonlinear Schrödinger equations in the mean-field limit regime for systems involving three-body terms. The inclusion of three-body interactions enriches the theory of quantum many-body systems, providing new insights into stabilization mechanisms, collective behaviors, and dynamical properties. Unlike conventional two-body interactions, which have been extensively studied in the BEC context, three-body terms introduce a higher-order nonlinearity capable of balancing attractive forces, preventing collapse, and stabilizing condensates in certain regimes, such as the supersolid phase \cite{BisBla-15,Blakie-16}. More recently, \cite{NguRic-24,NguRic-25} investigated quantum systems with both two-body and three-body interactions in the presence of an external trapping potential, obtaining results on mass concentration phenomena. It is a particularly interesting case that, for a Bose gas with an attractive two-body interaction and a repulsive three-body interaction, the system is self-trapped by the two-body attraction \cite{Petrov-14}. This happens in the low-dimensional space, i.e., $d \leq 2$, and this leads to the homogeneous BECs in the whole space. The present article addresses this phenomenon in the framework of effective theory. In particular, by introducing the three-2D-body repulsion, the two-2D-body attraction has to be sufficiently negative.

In the whole $d$ ($d\leq 2$) dimensional space, we consider the minimization problem
\begin{equation}\label{energy:NLS}
E_{a,b}^{\rm NLS} := \inf\left\{\mathcal{E}^{\rm NLS}_{a,b}[u]: u\in H^1(\mathbb{R}^{d}), \|u\|_{L^{2}}^{2} = 1\right\}
\end{equation}
where $a,b>0$, and
\begin{equation}\label{functional:NLS}
\mathcal{E}_{a,b}^{\rm NLS}[u] := \int_{\mathbb{R}^{d}}\left[ |\nabla u|^{2} - \frac{a}{2} |u|^{4} + \frac{b}{6} |u|^{6} \right].
\end{equation}
The functional \eqref{functional:NLS} is commonly referred to as the cubic-quintic nonlinear Schrödinger (NLS) functional. The effective statistics parameters $a$ and $b$ are a key quantity in this setup, controlling the balance between the two-body and three-body interactions. In the underlying Bose gas, $a$ is typically proportional to the two-body scattering length multiplied by the particle density, whereas $b$ is related to effective three-body interaction coefficients, which themselves depend on the density and the microscopic interaction potential. These parameters therefore encode how the strength and range of interactions scale with the density, and they directly affect collective quantum properties such as stability, condensate depletion, and excitation spectra. When the system \eqref{functional:NLS} is subject to a trapping potential, the validity of \eqref{energy:NLS} as an effective description of the many-body quantum system has been rigorously established in \cite{NguRic-24, NguRic-25} (see also \cite{NamRicTri-23} for the purely repulsive case).

In the one-dimensional space, we note that \cite{NguRic-24} deal with attractive three-1D-body interactions, which cannot be too negative. But the result in \cite{NguRic-24} also holds for three-1D-body repulsion. Furthermore, it is worth noting that the two-1D-body interaction considered in \cite{NguRic-24} can be repulsive and it compensates for the (critical) attractive three-1D-body interaction. However, in the homogeneous case, the two-1D-body interaction has to be attractive since otherwise the NLS energy is trivial.

In the two-dimensional space, an attractive two-$\mathbb{ R}^{2}$-body interaction alone, without any counteracting force, would lead to collapse in $\mathbb{R}^2$ \cite{BraSacTolHul-95,BraSacHul-97,SacStoHul-98}. This is due to the fact that there is a balance between the kinetic energy and the two-body interaction. Then the maximum value for the strength of two-body interaction, namely $a_{*}$, is determined naturally by the optimal constant in the Gagliardo--Nirenberg inequality
\begin{equation}\label{ineq:gn}
\frac{a_{*}}{2} \|u\|_{L^{4}}^{4} \leq \|\nabla u\|_{L^{2}}^{2} \|u\|_{L^{2}}^{2},\quad \forall\, u\in H^1(\mathbb{R}^{2}).
\end{equation}
It is well-known (see \cite{Weinstein-83}) that \eqref{ineq:gn} has a unique optimizer (up to translation and dilation) which is positive radially symmetric decreasing. Furthermore, one can calculate numerically the constant $a_{*}$ which is approximated by $4\pi \times 0.93$ (see e.g., \cite{SulSulPat-84}). It is interesting that adding a three-2D-body repulsive interaction introduces a stabilizing effect \cite{EsrGreZhoLin-96,JosRic-97,GamFreTom-99,AkhDasVag-99,GamFreTomCho-00}. This term penalizes high densities and provides a counteracting force against the clustering tendency driven by the two-2D-body attraction. Specifically, the repulsive three-2D-body interaction grows faster than the attractive two-2D-body interaction as the density increases. This differential growth rate ensures that no matter how large the attractive term becomes (i.e., even if the coefficient $a$ of the attraction term is arbitrarily large), the repulsive term can always compensate. In other words, the three-2D-body repulsion provides a sufficiently strong stabilization against an unbounded attraction. We refer readers to \cite{NguRic-25} for a mathematically rigorous proof of the stabilization against collapse of a trapped Bose gas.

In the analysis of trapped BECs, particularly in the study of Bose gases and other large condensate systems, the Thomas--Fermi (TF) approximation is commonly employed \cite{DalGioPitStr-99,SchVin-00}. To the best of our knowledge, our work provides the first rigorous derivation of a TF-type functional in the context of Bose gases with repulsive three-body interactions in low dimensions. This approximation is based on the assumption that, in the large effective statistics limit, the kinetic energy term in the focusing system when three-body repulsion is considered, can be neglected compared to the nonlinear terms. This is useful for understanding properties such as density profiles and phase transitions in a homogeneous Bose gas. In the viewpoint of the nonlinear theory \eqref{functional:NLS}, this phenomenon happens when the mass-critical nonlinearity (i.e., a nonlinearity whose scaling matches that of the kinetic term) is too strong. Mathematically, the TF regime corresponds to the limit $b \to +\infty$ in the one-dimensional case, and the limit $a \to +\infty$ in the two-dimensional case. Without the kinetic term in \eqref{functional:NLS}, the minimization problem \eqref{energy:NLS} is exactly soluble and gives a density profile of the Thomas--Fermi-type
\begin{equation}\label{energy:TF}
E^{\rm TF}_{a,b} = \inf\left\{\mathcal{E}^{\rm TF}_{a,b}[\varrho] : 0\leq \varrho \in L^{1} \cap L^{3}(\mathbb R^{d}), \int_{\mathbb R^{d}}\varrho = 1\right\}
\end{equation}
where
\begin{equation}\label{functional:TF}
\mathcal{E}^{\rm TF}_{a,b}[\varrho] = \int_{\mathbb R^{d}} \left[-\frac{a}{2}\varrho^{2} + \frac{b}{6}\varrho^{3}\right].
\end{equation}
We shall show that the minimization problem \eqref{energy:TF} admits (up to translation) a unique minimizer $\varrho^{\rm TF}_{a,b}$, which is positive radially symmetric decreasing. We also calculate explicitly the TF minimizer. Note that, by a simple scaling, we have
\begin{equation}\label{energy:TF-scaling}
E^{\rm TF}_{a,b} = \frac{a^{2}}{b}E^{\rm TF}_{1,1} \quad \text{and} \quad \varrho^{\rm TF}_{a,b}(x) = \frac{a}{b}\varrho^{\rm TF}_{1,1}\left(\left(\frac{a}{b}\right)^{\frac{1}{d}}x\right)
\end{equation}
where $\varrho^{\rm TF}_{1,1}$ is a positive radially symmetric decreasing minimizer for $E^{\rm TF}_{1,1}$.

Our first result concerns the existence of NLS and TF minimizer(s). We have the following.

\begin{theorem}\label{thm:existence}
Let $d \leq 2$. We have the following.
\begin{enumerate}[label=(\roman*)]
\item\label{thm:existence-2d} {\bf Existence of NLS minimizers.} The minimization problem $E_{a,b}^{\rm NLS}$ given by \eqref{energy:NLS} admits minimizers for every fixed $a > 0$ and $b \geq 0$ if $d=1$, and for every fixed $a>a_{*}$ and $b>0$ if $d=2$. Furthermore, $E_{a,b}^{\rm NLS} < 0$.

\item {\bf Existence and uniqueness of TF minimizer.} The minimization problem $E_{a,b}^{\rm TF}$ given by \eqref{energy:TF} admits a (unique) minimizer for every fixed $a,b>0$. Furthermore, $E_{a,b}^{\rm TF} = -\dfrac{3a^{2}}{8b}$.
\end{enumerate}
\end{theorem}

In much of the existing literature (see, e.g., \cite{GuoSei-14}), stabilization is achieved through an external (trapping) potential rather than through three-body interactions. In the present work, the latter play an analogous role, providing the confining effect needed to prevent collapse and ensure the existence of stable states. In \cite{NguRic-25}, the authors studied a setting that includes both a three-body interaction term and an external trapping potential, showing how these two mechanisms jointly stabilize the system and influence the condensate’s ground-state structure. It is a novelty that this stabilization can be obtained without external potentials, as shown in \Cref{thm:existence}.

Next, we investigate the mass concentration phenomenon for a homogeneous BEC with attractive two-body interactions and repulsive three-body interactions, within the framework of effective models. Our next result is as follows.

\begin{theorem}\label{thm:behavior}
Let $d \leq 2$ and $\{a_{n}\},\{b_{n}\} \subset (0, \infty)$ be such that $a_{n} \to a_{0} \in (0,\infty]$ and $b_{n} \to b_{0} \in [0,\infty]$ if $d=1$, and $a_{n} \to a_{0} \in [a_{*},\infty]$ and $b_{n} \to b_{0} \in [0,\infty]$ if $d=2$.

Let $\{u_{n}\}$ be a sequence of minimizers of $E_{a_{n}, b_{n}}^{\rm NLS}$ given by \eqref{energy:NLS}. We have the following.

\begin{enumerate}[label=(\roman*)]
\item {\bf Thomas--Fermi approximation.} Assume that $b_{0} = +\infty$ if $d=1$, and $a_{0} = +\infty$ if $d=2$. Then, up to a translation,
\begin{equation}\label{cv:minimizer-TF}
\lim_{n\to+\infty} \frac{b_{n}}{a_{n}} u_{n}\left(\left(\frac{b_{n}}{a_{n}}\right)^{\frac{1}{d}} \cdot\right)^{2} = \varrho^{\rm TF}_{1,1}
\end{equation}
strongly in $L^{1}\cap L^{3}(\mathbb{R}^{d})$. Furthermore,
\begin{equation}\label{cv:energy-TF}
\lim_{n\to\infty}\frac{b_{n}}{a_{n}^{2}}E_{a_{n}, b_{n}}^{\rm NLS} = E^{\rm TF}_{1,1}.
\end{equation}

\item {\bf 1D cubic(-quintic) approximation.} In the one-dimensional case $d=1$, we assume that $0 \leq b_{0} < \infty$. Then, up to a translation and extracting a subsequence,
\begin{equation}\label{cv:minimizer-cubic-quintic-1d}
\lim_{n\to\infty}a_{n}^{-\frac{1}{2}}u_{n}(a_{n}^{-1}\cdot) = u_{0}
\end{equation}
strongly in $H^{1}(\mathbb R)$, where $u_{0}$ is a minimizer for $E^{\rm NLS}_{1,b_{0}}$. Furthermore,
\begin{equation}\label{cv:energy-cubic-quintic-1d}
\lim_{n\to\infty} a_{n}^{-2}E^{\rm NLS}_{a_{n},b_{n}} = E^{\rm NLS}_{1,b_{0}}.
\end{equation}

\item {\bf 2D cubic(-quintic) approximation.} In the two-dimensional case $d=2$, we distinguish two cases.
\begin{itemize}
\item If $a_{*} < a_{0} < \infty$ then, up to a translation and extracting a subsequence,
\begin{equation}\label{cv:minimizer-cubic-quintic-2d}
\lim_{n\to\infty} b_{n}^{\frac{1}{2}}u_{n}(b_{n}^{\frac{1}{2}}\cdot) = u_{0}
\end{equation}
strongly in $H^{1}(\mathbb R^{2})$, where $u_{0}$ is a minimizer for $E^{\rm NLS}_{a_{0},1}$. Furthermore,
\begin{equation}\label{cv:energy-cubic-quintic-2d}
\lim_{n\to\infty} b_{n}E^{\rm NLS}_{a_{n},b_{n}} = E^{\rm NLS}_{a_{0},1}.
\end{equation}

\item If $a_{0} = a_{*}$ then, up to a translation,
\begin{equation}\label{cv:minimizer-cubic-2d}
\lim_{n\to+\infty} \ell_{n}^{\frac{1}{2}} u_{n}(\ell_{n}^{\frac{1}{2}} \cdot) = Q_0 \quad \text{with} \quad \ell_{n} = \frac{2\|Q_{0}\|_{L^{6}}^{6}b_{n}}{3\|Q_{0}\|_{L^{4}}^{4}(a_{n} - a_{*})}
\end{equation}
strongly in $H^1(\mathbb{R}^{2})$ for the whole sequence. Here $Q_{0}$ is the unique $L^{2}$-normalized optimizer for \eqref{ineq:gn}. Furthermore,
\begin{equation}\label{cv:energy-cubic-2d}
\lim_{n\to\infty}\frac{b_{n}}{(a_{n}-a_{*})^{2}}E^{\rm NLS}_{a_{n}, b_{n}} = -\frac{3}{8}\frac{\|Q_{0}\|_{L^{4}}^{8}}{\|Q_{0}\|_{L^{6}}^{6}}.
\end{equation}
\end{itemize}
\end{enumerate}
\end{theorem}

\begin{remark}
\begin{itemize}
\item A minimizer of $E_{a,b}^{\rm NLS}$ (if it exists) is nonnegative, radially symmetric, and decreasing, by rearrangement inequalities (see, e.g., \cite{LieLos-01}). Furthermore, both $E_{a,b}^{\rm NLS}$ and $E_{a,b}^{\rm TF}$ are strictly negative, for fixed $a,b>0$. Those are key properties in order to establish the existence of NLS/TF minimizers.
\item It leaves an open problem to study the homogeneous Bose gases in high-dimensional space ($d > 2$). In which case, the three-body repulsions always compensate the two-body interactions. The existence and uniqueness of TF minimizer in \Cref{thm:existence} are still valid, but the one of NLS minimizers is not yet clear. We hope to revisit this problem in the future.
\end{itemize}
\end{remark}

\Cref{thm:behavior} shows that the Thomas--Fermi-type energy functional \eqref{functional:TF} provides a leading-order algebraic description of the condensate density in regimes where the interaction energy dominates the kinetic energy. Since the TF functional contains no kinetic term, it yields explicit local relations for the density and is considerably simpler to analyze than the full cubic-quintic NLS functional. This simplicity makes it particularly suitable for deriving scaling laws and bulk energy asymptotics. Rigorous derivations and discussions of TF limits for trapped BECs, as well as analyses of their validity, can be found in the literature (see, e.g., \cite{LieSeiYng-00} and subsequent works). The scaling considered here for self-trapped BECs is chosen precisely to place the system in this interaction-dominated regime, so that the TF functional provides the appropriate leading approximation, while the cubic-quintic NLS retains the subleading kinetic corrections.

\medskip
\noindent{\bf Acknowledgement.} This research was funded by Vietnam Ministry of Education and Training under grant number B2026-CTT-10.

\section{Existence of minimizers}

\subsection{Existence of NLS minimizers}\label{sec:existence-nls}

We first observe that $E^{\rm NLS}_{a,b} < 0$ for every fixed $a>0$ and $b \geq 0$ if $d = 1$, and for every fixed $a>a_{*}$ and $b>0$ if $d = 2$. Indeed, by the variational principle and a simple scaling argument, we have
$$
E^{\rm NLS}_{a,b} \leq \mathcal{E}^{\rm NLS}_{a,b}\left[\ell^{\frac{d}{2}} \phi(\ell\cdot)\right] = \int_{\mathbb{ R}^{d}} \left[\ell^{2}|\nabla \phi|^{2} - \ell^{d}\frac{a}{2}|\phi|^{4} + \ell^{2d} \frac{b}{6} |\phi|^{6} \right]
$$
where $\ell>0$ and $\phi \in H^{1}(\mathbb{R}^{d})$ is $L^{2}$-normalized test function chosen appropriately. The scaling relies on the standard dilation $\phi(x)\mapsto \ell^{d/2}\phi(\ell x)$, which preserves the mass, i.e., $\|\ell^{\frac{d}{2}} \phi(\ell\cdot)\|_{L^{2}} = \|\phi\|_{L^{2}} = 1$. In the case $d = 1$, for any fixed smooth function $\phi$, there exists $\ell = \ell(a,b,\phi) > 0$ sufficiently small such that $E^{\mathrm{NLS}}_{a,b} < 0$. In the case $d = 2$, if $\phi \equiv Q_{0}$ be the unique $L^{2}(\mathbb{R}^{2})$-normalized optimizer of \eqref{ineq:gn}, then
$$
E^{\rm NLS}_{a,b} \leq \int_{\mathbb{ R}^{2}} \left[\ell^{2}\frac{a_{*}-a}{2}|Q_{0}|^{4} + \ell^{4} \frac{b}{6} |Q_{0}|^{6} \right] < 0.
$$
Here we have used the fact that $a > a_{*}$ and chose $\ell = \ell(a,b,Q_{0}) > 0$ sufficiently small.

We are now in the position to prove the existence of NLS minimizers by the method of variational calculus. Let $\{u_{n}\}$ be a minimizing sequence for $E^{\rm NLS}_{a,b}$, i.e., $\|u_{n}\|_{L^{2}}^{2} = 1$ and 
$$
E^{\rm NLS}_{a,b} = \lim_{n\to\infty}\mathcal{E}^{\rm NLS}_{a,b}[u_{n}].
$$
By rearrangement inequality, $\{u_{n}\}$ can be chosen radially symmetric decreasing. On the other hand, by H\"older inequality, we have
$$
\frac{a}{2} \int_{\mathbb R^{d}}|u_{n}|^{4} \leq \frac{b}{6} \int_{\mathbb R^{d}}|u_{n}|^{6} + C(a,b)\int_{\mathbb R^{d}}|u_{n}|^{2}.
$$
This implies that $\{u_{n}\}$ is uniformly bounded in $H^{1}(\mathbb{R}^{d})$. 
Up to a translation and extracting a subsequence, $u_{n} \to u_{0}$ weakly in $H^{1}(\mathbb R^{d})$ and pointwise almost everywhere in $\mathbb R^{d}$. On the other hand, by Sobolev embedding, $\{u_{n}\}$ is also bounded uniformly in $L^{p}(\mathbb R^{d})$, for all $2 \leq p \leq \infty$ if $d=1$, and for all $2 \leq p < \infty$ if $d=2$. In particular, since $\{u_{n}\} \subset L^{2}\cap L^{p_{0}}(\mathbb R^{d})$, for $d\leq 2$ and for any $2 < p_{0} < \infty$, is positive radially symmetric decreasing, we deduce that
$$
|u_{n}(x)| \leq |\{y:|y|<|x|\}|^{-\frac{1}{p}} \left(\int_{|y|<|x|} |u_{n}(y)|^{p} {\rm d}y\right)^{\frac{1}{p}} \leq |B_{\mathbb R^{d}}(0,1)|^{-\frac{1}{p}}|x|^{-\frac{d}{p}} \|u_{n}\|_{L^{p}}
$$
for all $x \in \mathbb{R}^{d}$ (with $d \leq 2$) and for every $2 \leq p \leq p_{0}$. We then deduce the following uniform bound in $n$,
$$
|u_{n}(x)| \lesssim \min \left\{|x|^{-\frac{d}{2}},|x|^{-\frac{d}{p_{0}}}\right\}.
$$
Let's denote by $f$ the right-hand side of the above. By direct calculation, we can check that, for every $2<q<p_{0}$,
$$
\int_{\mathbb R^{d}}f(x)^{q}{\rm d}x \leq \int_{|x| \leq 1} |x|^{-\frac{dq}{p_{0}}}{\rm d}x + \int_{|x| \geq 1} |x|^{-\frac{dq}{2}}{\rm d}x < \infty.
$$
By Dominated Convergence Theorem, we have that $u_{n}$ converges strongly to $u_{0}$ in $L^{q}(\mathbb R^{d})$, for every $2<q<\infty$.

We are now in the position to show that $u_{0}$ is actually a minimizer for $E^{\rm NLS}_{a,b}$. We first claim that $u_{0} \not\equiv 0$ since otherwise we must have that
$$
E^{\rm NLS}_{a,b} = \lim_{n\to\infty}\mathcal{E}^{\rm NLS}_{a,b}[u_{n}] \geq 0,
$$
where we have used the strong convergence $u_{n}$ to $0$ in $L^{4}(\mathbb R^{d})$ and the nonnegativity of the kinetic and quintic terms. This contradicts the fact that $E^{\rm NLS}_{a,b} < 0$. At this stage, we use Fatou's lemma, the weak convergence $u_{n}$ to $u_{0}$ in $H^{1}(\mathbb R^{d})$, as well as the strong convergence $L^{4}\cap L^{6}(\mathbb R^{d})$ in order to obtain that
\begin{align*}
E^{\rm NLS}_{a,b} = \lim_{n\to\infty}\mathcal{E}^{\rm NLS}_{a,b}[u_{n}] \geq \mathcal{E}^{\rm NLS}_{a,b}[u_{0}] & = \int_{\mathbb{R}^{d}}\left[ \|u_{0}\|_{L^{2}}^{2-\frac{4}{d}}|\nabla \widetilde{u_{0}}|^{2} - \frac{a}{2} \|u_{0}\|_{L^{2}}^{2} |\widetilde{u_{0}}|^{4} + \frac{b}{6} \|u_{0}\|_{L^{2}}^{2} |\widetilde{u_{0}}|^{6} \right] \\
& \geq \|u_{0}\|_{L^{2}}^{2} \mathcal{E}^{\rm NLS}_{a,b}[\widetilde{u_{0}}] \\
& \geq \|u_{0}\|_{L^{2}}^{2} E^{\rm NLS}_{a,b} \\
& \geq E^{\rm NLS}_{a,b}.
\end{align*}
Here we have set $\widetilde{u_{0}} = u_{0}\big(\|u_{0}\|_{L^{2}}^{\frac{2}{d}}\cdot\big)$ with $1 \leq d \leq 2$, and we have used the facts that $0< \|u_{0}\|_{L^{2}}^{2} \leq 1$ (by Fatou's lemma and the weak convergence $u_{n} \wto u_{0}$ in $L^{2}(\mathbb R^{d})$) and that $E^{\rm NLS}_{a,b} < 0$. The equality must occur in the above and we must have that $\|u_{0}\|_{L^{2}}^{2} = 1$. This also proves that $u_{0}$ is a minimizer of $E^{\rm NLS}_{a,b}$.

\subsection{Existence and uniqueness of Thomas--Fermi minimizer}

In this subsection, we prove the existence and uniqueness of TF minimizer. We also calculate explicitly the Thomas--Fermi energy. For simplicity, we compute $E^{\rm TF}_{1,1}$ and the other values of $E^{\rm TF}_{a,b}$ for $a,b>0$ can be calculated by a scaling argument \eqref{energy:TF-scaling}. On the one hand, we use Cauchy--Schwarz inequality to estimate the energy from below as follows
$$
\mathcal{E}^{\rm TF}_{a,b}[\varrho] \geq -\frac{3}{8}\int_{\mathbb R^{d}}\varrho = -\frac{3}{8}
$$
for every $0 \leq \varrho \in L^{1} \cap L^{3}(\mathbb R^{d})$ with $\int_{\mathbb R^{d}}\varrho = 1$. This yields that $E^{\rm TF}_{1,1}$ is at least $-\frac{3}{8}$. In addition, $E^{\rm TF}_{1,1}$ is strictly negative. Indeed, by the variational principle and a simple scaling argument, we have 
$$
E^{\rm TF}_{1,1} \leq \mathcal{E}^{\rm TF}_{1,1}[\left[\ell^{d} \phi(\ell\cdot)\right]] = \int_{\mathbb R^{d}} \left[-\ell^{d}\frac{a}{2}\phi^{2} + \ell^{2d}\frac{b}{6}\phi^{3}\right] < 0,
$$
where $\phi$ is a fixed nonnegative $L^{1}(\mathbb R^{d})$ smooth function such that $\int_{\mathbb R^{d}}\phi = 1$, and we have chosen $\ell = \ell(a,b,\phi) > 0$ sufficient small. On the other hand, the existence of a minimizer $\varrho_{1,1}$ for $E^{\rm TF}_{1,1}$ can be established by the same arguments as in \Cref{sec:existence-nls}. For the reader’s convenience, we recall that any minimizing sequence for $E^{\rm TF}_{1,1}$ is uniformly bounded in $L^{1} \cap L^{3}(\mathbb{R}^{d})$ and can be taken to be radially symmetric. Consequently, it converges strongly in $L^{p}(\mathbb{R}^{d})$ for every $1 < p < 3$. This compactness property together with the negativity of the TF energy then be used, via the direct method in the calculus of variations, to prove the existence of a TF minimizer (see \Cref{sec:tf} for similar arguments). Such a minimizer satisfies the Euler--Lagrange equation
$$
\frac{\varrho_{1,1}^{2}}{2} = \left[\varrho_{1,1} + \lambda\right]_{+}
$$
where $\lambda \in \mathbb{ R}$ is the Lagrange multiplier. The above equation is solvable and the positive solution is a constant. In order to satisfy the constraint of unit mass, it is of the form
$$
\varrho_{1,1} = |E|^{-1}\mathbbm{1}_{E}
$$
where $E$ is a finite Lebesgue measure set.
We use the above function as a trial state for $E^{\rm TF}_{1,1}$ and obtain that
$$
E^{\rm TF}_{1,1} \leq \mathcal{E}^{\rm TF}_{a,b}[\varrho_{1,1}] = -\frac{1}{2|E|} + \frac{1}{6|E|^{2}} = -\frac{3}{8},
$$
where we have chosen $E$ such that $|E| = \dfrac{2}{3}$.

\section{Local density approximation and other limit regimes}

\subsection{Thomas--Fermi approximation}\label{sec:tf}

We show that the kinetic term can be neglected in the regime where $b_{n} \to +\infty$ in the one-dimensional case, and $a_{n} \to + \infty$ in the two-dimensional case. Therefore, the NLS theory is well approximated by the Thomas--Fermi theory. We prove \eqref{cv:minimizer-TF} and \eqref{cv:energy-TF}.

Obviously, $E^{\rm NLS}_{a_{n},b_{n}} \geq E^{\rm TF}_{a_{n},b_{n}}$, for every $a_{n},b_{n}>0$, so the energy lower bound in \eqref{cv:energy-TF} is immediate. In order to prove the matching energy upper bound in \eqref{cv:energy-TF}, we consider 
$$
u_{n} = \sqrt{g_{\mu}*\varrho^{\rm TF}_{n}} \quad \text{with} \quad \varrho^{\rm TF}_{n}(x) = \frac{a_{n}}{b_{n}}\varrho^{\rm TF}_{1,1}\left(\left(\frac{a_{n}}{b_{n}}\right)^{\frac{1}{d}}x\right)
$$
as the trial state for $E^{\rm NLS}_{a_{n},b_{n}}$. Here $\varrho^{\rm TF}_{1,1}$ denotes the (unique) minimizer for $E^{\rm TF}_{1,1}$, and
$$
g_{\mu} = \dfrac{\Gamma(\frac{d}{2})}{2\pi^{\frac{d}{2}}\Gamma(d)}\mu^{\frac{d}{2}}e^{-\sqrt{\mu} |x|},
$$
with $d \leq 2$ and $\mu>0$ to be chosen. It is worth noting that $\int_{\mathbb R^{d}}g_{\mu} = 1$ and $|\nabla g_{\mu}| = \sqrt{\mu} g_{\mu}$, for all $\mu>0$. By the variational principle, we have
\begin{align*}
E^{\rm NLS}_{a_{n},b_{n}} & \leq \mathcal{E}^{\rm NLS}_{a_{n},b_{n}}[u_{n}] \\
& = \int_{\mathbb R^{d}} \frac{|(\nabla g_{\mu}*\varrho^{\rm TF}_{n})(x)|^{2}}{4(g_{\mu}*\varrho^{\rm TF}_{n})(x)}{\rm d}x - \frac{a_{n}}{2} \int_{\mathbb R^{d}}(g_{\mu}*\varrho^{\rm TF}_{n})(x)^{2} + \frac{b_{n}}{6} \int_{\mathbb R^{d}}(g_{\mu}*\varrho^{\rm TF}_{n})(x)^{3} \\
& \leq \frac{\mu}{4} + \frac{a_{n}}{2}\left(\int_{\mathbb R^{d}}\varrho^{\rm TF}_{n}(x)^{2} - \int_{\mathbb R^{d}}(g_{\mu}*\varrho^{\rm TF}_{n})(x)^{2}\right) + E^{\rm TF}_{a_{n},b_{n}} \\
& = \frac{\mu}{4} + \frac{a_{n}^{2}}{2b_{n}}\left(\int_{\mathbb R^{d}}\varrho^{\rm TF}_{1,1}(x)^{2} - \int_{\mathbb R^{d}}\left(g_{\mu \left(\frac{a_{n}}{b_{n}}\right)^{-\frac{2}{d}}}*\varrho^{\rm TF}_{1,1}\right)(x)^{2}\right) + E^{\rm TF}_{a_{n},b_{n}},
\end{align*}
where we have used Young's inequality for the three-body term. Now, we choose $\mu = \mu_{n}$ in such a way that, either when $b_{n} \to +\infty$ for $d=1$ or when $a_{n} \to + \infty$ for $d=2$,
$$
\mu = \mu_{n} = o\left(E^{\rm TF}_{a_{n},b_{n}}\right) = o\left(\frac{a_{n}^{2}}{b_{n}}\right) \quad \text{and} \quad \mu \left(\frac{a_{n}}{b_{n}}\right)^{-\frac{2}{d}} \to +\infty.
$$
For instance, we can take
$$
\mu = \frac{a_{n}^{1+\frac{1}{d}}}{b_{n}^{\frac{1}{2}+\frac{1}{d}}}.
$$
Consequently, we have the following strong convergence in $L^{2}(\mathbb R^{d})$ (see, e.g., \cite[Theorem 2.16]{LieLos-01})
$$
\lim_{n\to\infty}g_{\mu \left(\frac{a_{n}}{b_{n}}\right)^{-\frac{2}{d}}}*\varrho^{\rm TF}_{1,1} = \varrho^{\rm TF}_{1,1}.
$$
Putting all together, we obtain the desired energy upper bound in \eqref{cv:energy-TF}.

Next, we prove \eqref{cv:minimizer-TF}. Let $\{u_{n}\}$ be a sequence of minimizers for $E^{\rm NLS}_{a_{n},b_{n}}$. Setting 
$$
\varrho_{n}(x) = \frac{b_{n}}{a_{n}} u_{n}\left(\left(\frac{b_{n}}{a_{n}}\right)^{\frac{1}{d}}x\right)^{2}.
$$
Obviously, by neglecting the nonnegative kinetic term, one can estimate
$$
E^{\rm NLS}_{a_{n},b_{n}} \geq \int_{\mathbb R^{d}} \left[ -\frac{a_{n}}{2}|u_{n}|^{4} + \frac{b_{n}}{6}|u_{n}|^{6}\right] = \frac{a_{n}^{2}}{b_{n}} \int_{\mathbb R^{d}} \left[ -\frac{1}{2}\varrho_{n}^{2} + \frac{1}{6}\varrho_{n}^{3}\right].
$$
This, together with the energy upper bound in \eqref{cv:energy-TF}, implies that
\begin{equation}\label{cv:energy-TF-lower-bound}
E^{\rm TF}_{1,1} \geq \int_{\mathbb R^{d}} \left[ -\frac{1}{2}\varrho_{n}^{2} + \frac{1}{6}\varrho_{n}^{3}\right].
\end{equation}
By the H\"older inequality, we deduce from the above that $\{\varrho_{n}\}_{n}$ is bounded uniformly in $L^{1}\cap L^{3}(\mathbb R^{d})$. Up to translation and extracting a subsequence, $\varrho_{n} \to \varrho_{0}$ weakly in $L^{1}\cap L^{3}(\mathbb R^{d})$. We show that this is actually the strong convergence in $L^{1}\cap L^{3}(\mathbb R^{d})$. We observe that $\{\varrho_{n}\} \subset L^{1}\cap L^{3}(\mathbb R^{d})$ is positive, radially symmetric, and decreasing because $\{u_{n}\}$ is. By the same arguments as in the proof of \Cref{thm:existence}, we can show that $\varrho_{n}$ converges strongly to $\varrho_{0}$ in $L^{q}(\mathbb R^{d})$, for every $1<q<3$, and pointwise almost everywhere in $\mathbb R^{d}$. We prove that this convergence also holds true in $L^{1}(\mathbb R^{d})$ and $L^{3}(\mathbb R^{d})$ as well, which yields in particular that $\varrho_{0}$ is a minimizer for $E^{\rm TF}_{1,1}$.

We first note that the limiting profile $\varrho_{0}$ is non-trivial since otherwise we must have that $E^{\rm TF}_{1,1}$ is nonnegative, by taking the limit $n\to\infty$ in \eqref{cv:energy-TF-lower-bound}. This contradicts the negativity of $E^{\rm TF}_{1,1}$. On the other hand, taking again the limit $n \to \infty$ in \eqref{cv:energy-TF-lower-bound}, using Fatou's lemma and the weak convergence $\varrho_{n} \wto \varrho_{0}$ in $L^{3}(\mathbb R^{d})$ as well as the strong convergence $\varrho_{n} \to \varrho_{0}$ in $L^{2}(\mathbb R^{d})$, we obtain
$$
E^{\rm TF}_{1,1} \geq \int_{\mathbb R^{d}} \left[ -\frac{1}{2}\varrho_{0}^{2} + \frac{1}{6}\varrho_{0}^{3}\right] = \int_{\mathbb R^{d}} \left[ -\frac{1}{2}\widetilde{\varrho_{0}}^{2} + \frac{1}{6}\widetilde{\varrho_{0}}^{3}\right] \int_{\mathbb R^{d}}\varrho_{0} \geq E^{\rm TF}_{1,1} \int_{\mathbb R^{d}}\varrho_{0} \geq E^{\rm TF}_{1,1}.
$$
Here we denote $\widetilde{\varrho_{0}}(x) = \varrho_{0}\left(\big(\int_{\mathbb R^{d}}\varrho_{0} \big)^{\frac{1}{d}}x\right)$, which satisfies $\int_{\mathbb R^{d}}\widetilde{\varrho_{0}} = 1$, and we have used the facts that $0 < \int_{\mathbb R^{d}}\varrho_{0} \leq 1$ (by Fatou's lemma and the weak convergence $\varrho_{n} \wto \varrho_{0}$ in $L^{1}(\mathbb R^{d})$) and that $E^{\rm TF}_{1,1} < 0$.

The equality must occur in the above and we must have that $\int_{\mathbb R^{d}}\varrho_{0} = 1$ and that $\int_{\mathbb R^{d}}\varrho_{n}^{3} \to \int_{\mathbb R^{d}}\varrho_{0}^{3}$. By Br\'ezis--Lieb lemma \cite{BreLie-83}, $\varrho_{n} \to \varrho_{0}$ strongly in $L^{1} \cap L^{3}(\mathbb R^{d})$. This also proves that $\varrho_{0}$ is a minimizer for $E^{\rm TF}_{1,1}$. The proof is completed.

\subsection{1D cubic(-quintic) approximation} In this subsection, we consider the 1D cubic-quintic in the regime other than Thomas--Fermi, i.e., $b_{0} \ne \infty$. In particular, the three-body interaction vanishes in the case $b_{0} \equiv 0$ and the 1D cubic-quintic problem reduces to the 1D cubic problem.

Let $u_{0}$ be a minimizer for $E^{\rm NLS}_{1,b_{0}}$. By the variational principle,
$$
E^{\rm NLS}_{a_{n},b_{n}} \leq \mathcal{E}^{\rm NLS}_{a_{n},b_{n}}\left[a_{n}^{\frac{1}{2}}u_{0}(a_{n}\cdot)\right] = a_{n}^{2}\mathcal{E}^{\rm NLS}_{1,b_{n}}[u_{0}].
$$
Dividing both sides by $a_{n}^{2}$ and taking the limit $n\to\infty$, we thus obtain the energy upper bound in \eqref{cv:energy-cubic-quintic-1d}, i.e.,
\begin{equation}\label{cv:energy-cubic-quintic-1d-upper-bound}
\lim_{n\to\infty} a_{n}^{-2}E^{\rm NLS}_{a_{n},b_{n}} \leq \mathcal{E}^{\rm NLS}_{1,b_{0}}[u_{0}] = E^{\rm NLS}_{1,b_{0}}.
\end{equation}
We prove the matching energy lower bound in \eqref{cv:energy-cubic-quintic-1d} by showing the convergence of minimizers in \eqref{cv:minimizer-cubic-quintic-1d}. For this purpose, we set
$$
w_{n}(x) := a_{n}^{-\frac{1}{2}}u_{n}(a_{n}^{-1}x).
$$
We rewrite the energy functional as follows
\begin{equation}\label{cv:energy-cubic-quintic-1d-lower-bound}
E_{a_{n},b_{n}}^{\rm NLS} = \mathcal{E}_{a_{n},b_{n}}^{\rm NLS}[u_{n}] = a_{n}^{2}\mathcal{E}^{\rm NLS}_{1,b_{n}}[w_{n}].
\end{equation}
Since the two-body interaction is mass-subcritical with respect to the kinetic energy and the three-body interaction is non-negative, we deduce from \eqref{cv:energy-cubic-quintic-1d-upper-bound} and \eqref{cv:energy-cubic-quintic-1d-lower-bound} the uniform boundedness of $\{w_{n}\}$ in $H^{1}(\mathbb R)$. By Sobolev embedding, $\{w_{n}\}$ is also bounded uniformly in $L^{p}(\mathbb R^{2})$, for every $2\leq p \leq \infty$. Note that $\{w_{n}\}$ is radially symmetric decreasing since $\{u_{n}\}$ is. By the same argument in the proof of \Cref{thm:existence}, we have that (up to a translation and extracting a subsequence) $w_{n} \to w_{0}$ weakly in $H^{1}(\mathbb R)$, strongly in $L^{p}(\mathbb R)$ for every $2<p<\infty$, and pointwise almost everywhere in $\mathbb R$. We prove that this is really strong convergence in $H^{1}(\mathbb R)$. We first claim that $w_{0} \not\equiv 0$ since otherwise we must have, by \eqref{cv:energy-cubic-quintic-1d-upper-bound} and \eqref{cv:energy-cubic-quintic-1d-lower-bound},
$$
E^{\rm NLS}_{1,b_{0}} \geq 0.
$$
This contradicts the negativity of $E^{\rm NLS}_{1,b_{0}}$, for every $0 \leq b_{0} < \infty$. Now, we use again \eqref{cv:energy-cubic-quintic-1d-upper-bound}, \eqref{cv:energy-cubic-quintic-1d-lower-bound}, Fatou's lemma and the weak convergence $w_{n} \wto w_{0}$ in $H^{1}(\mathbb R)$ as well as the strong convergence $w_{n} \to w_{0}$ in $L^{4} \cap L^{6} (\mathbb R)$ to obtain that
$$
E^{\rm NLS}_{1,b_{0}} \geq \mathcal{E}^{\rm NLS}_{1,b_{0}}[w_{0}] = \|w_{0}\|_{L^{2}}^{2} \mathcal{E}^{\rm NLS}_{1,b_{0}}[\widetilde{w_{0}}] \geq \|w_{0}\|_{L^{2}}^{2} E^{\rm NLS}_{1,b_{0}} \geq E^{\rm NLS}_{1,b_{0}}.
$$
Here we have set $\widetilde{w_{0}} = w_{0}\big(\|w_{0}\|_{L^{2}}^{2}\cdot\big)$ and we have used that $0< \|w_{0}\|_{L^{2}}^{2} \leq 1$ (by Fatou's lemma and the weak convergence $w_{n} \wto w_{0}$ in $L^{2}(\mathbb R)$) as well as $E^{\rm NLS}_{1,b_{0}} < 0$, for every $0 \leq b_{0} < \infty$. The equality must occur in the above and we must have that $\|w_{0}\|_{L^{2}}^{2} = 1$ and that $\|\nabla w_{n}\|_{L^{2}}^{2} \to \|\nabla w_{0}\|_{L^{2}}^{2}$. By Br\'ezis--Lieb lemma, $w_{n} \to w_{0}$ strongly in $H^{1}(\mathbb R)$. This also proves that $w_{0}$ is a minimizer for $E^{\rm NLS}_{1,b_{0}}$. The proof is completed.

\subsection{2D cubic(-quintic) approximation} In this subsection, we consider the 2D cubic-quintic in the regime other than Thomas--Fermi, i.e., $a_{0} \ne \infty$. In which case, the two-body interaction does not vanish since $a_{0}$ varies between $a_{*}$ and infinity. We distinguish the two cases: either $a_{*} < a_{0} < \infty$ or $a_{0} = a_{*}$. In the former case, we obtain the cubic-quintic approximation which is similar to the 1D Bose gases.

We first prove \eqref{cv:minimizer-cubic-quintic-2d} and \eqref{cv:energy-cubic-quintic-2d}. Let $u_{0}$ be a minimizer for $E^{\rm NLS}_{a_{0},1}$. By the variational principle,
$$
E^{\rm NLS}_{a_{n},b_{n}} \leq \mathcal{E}^{\rm NLS}_{a_{n},b_{n}}\left[b_{n}^{-\frac{1}{2}}u_{0}\big(b_{n}^{-\frac{1}{2}}\cdot\big)\right] = b_{n}^{-1}\mathcal{E}^{\rm NLS}_{a_{n},1}[u_{0}].
$$
Multiplying both sides by $b_{n}$ and taking the limit $n\to\infty$, we thus obtain the energy upper bound in \eqref{cv:energy-cubic-quintic-1d}, i.e.,
\begin{equation}\label{cv:energy-cubic-quintic-2d-upper-bound}
\lim_{n\to\infty} b_{n}E^{\rm NLS}_{a_{n},b_{n}} \leq \mathcal{E}^{\rm NLS}_{a_{0},1}[u_{0}] = E^{\rm NLS}_{a_{0},1}.
\end{equation}
We prove the matching energy lower bound in \eqref{cv:energy-cubic-quintic-2d} by showing the convergence of minimizers in \eqref{cv:minimizer-cubic-quintic-2d}. For this purpose, we set
$$
w_{n}(x) := b_{n}^{\frac{1}{2}}u_{n}\left(b_{n}^{\frac{1}{2}}x\right).
$$
We rewrite the energy functional as follows
\begin{equation}\label{cv:energy-cubic-quintic-2d-lower-bound}
E_{a_{n},b_{n}}^{\rm NLS} = \mathcal{E}_{a_{n},b_{n}}^{\rm NLS}[u_{n}] = b_{n}^{-1}\mathcal{E}^{\rm NLS}_{a_{n},1}[w_{n}].
\end{equation}
Since $\{a_{n}\}$ is bounded and the attractive two-body interaction is controlled by the repulsive three-body interaction, by H\"older inequality, we deduce from \eqref{cv:energy-cubic-quintic-2d-upper-bound} and \eqref{cv:energy-cubic-quintic-2d-lower-bound} the uniform boundedness of $\{w_{n}\}$ in $H^{1}(\mathbb R^{2})$. By Sobolev embedding, $\{w_{n}\}$ is also bounded uniformly in $L^{p}(\mathbb R^{2})$, for every $2\leq p < \infty$. Note that $\{w_{n}\}$ is radially symmetric decreasing since $\{u_{n}\}$ is. By the same argument as in the proof of \Cref{thm:existence}, we have that (up to a translation and extracting a subsequence) $w_{n} \to w_{0}$   weakly in $H^{1}(\mathbb R^{2})$, strongly in $L^{p}(\mathbb R^{2})$ for every $2<p<\infty$, and pointwise almost everywhere in $\mathbb R^{2}$. We prove that this is really strong convergence in $H^{1}(\mathbb R^{2})$. We first claim that $w_{0} \not\equiv 0$ since otherwise we must have, by \eqref{cv:energy-cubic-quintic-2d-upper-bound} and \eqref{cv:energy-cubic-quintic-2d-lower-bound},
$$
E^{\rm NLS}_{a_{0},1} \geq 0.
$$
This contradicts the negativity of $E^{\rm NLS}_{a_{0},1}$, for every $a_{*} < a_{0} < \infty$. Now, we use again \eqref{cv:energy-cubic-quintic-2d-upper-bound}, \eqref{cv:energy-cubic-quintic-2d-lower-bound}, Fatou's lemma and the weak convergence $w_{n} \wto w_{0}$ in $H^{1}(\mathbb R^{2})$ as well as the strong convergence $w_{n} \to w_{0}$ in $L^{4} \cap L^{6} (\mathbb R^{2})$ to obtain that
$$
E^{\rm NLS}_{a_{0},1} \geq \mathcal{E}^{\rm NLS}_{a_{0},1}[w_{0}] = \mathcal{E}^{\rm NLS}_{\|w_{0}\|_{L^{2}}^{2}a_{0},\|w_{0}\|_{L^{2}}^{2}}[\widetilde{w_{0}}] \geq \|w_{0}\|_{L^{2}}^{2}\mathcal{E}^{\rm NLS}_{a_{0},1}[\widetilde{w_{0}}] \geq \|w_{0}\|_{L^{2}}^{2} E^{\rm NLS}_{a_{0},1} \geq E^{\rm NLS}_{a_{0},1}.
$$
Here we have set $\widetilde{w_{0}} = w_{0}\big(\|w_{0}\|_{L^{2}}\cdot\big)$ and we have used that $0< \|w_{0}\|_{L^{2}}^{2} \leq 1$ (by Fatou's lemma and the weak convergence $w_{n} \wto w_{0}$ in $L^{2}(\mathbb R^{2})$) as well as $E^{\rm NLS}_{a_{0},1} < 0$, for every $a_{*} < a_{0} < \infty$. The equality must occur in the above and we must have that $\|w_{0}\|_{L^{2}}^{2} = 1$ and that $\|\nabla w_{n}\|_{L^{2}}^{2} \to \|\nabla w_{0}\|_{L^{2}}^{2}$. By Br\'ezis--Lieb lemma, $w_{n} \to w_{0}$ strongly in $H^{1}(\mathbb R^{2})$. This also proves that $w_{0}$ is a minimizer for $E^{\rm NLS}_{a_{0},1}$. The proof is completed.

Finally, we consider the case where $a_{0} = a_{*}$. This corresponds to the limit $a_{n} \searrow a_{*}$. We prove \eqref{cv:minimizer-cubic-2d} and \eqref{cv:energy-cubic-2d}. Let $Q_{0}$ be the unique $L^{2}$-normalized optimizer for \eqref{ineq:gn}. By the variational principle,
\begin{align}\label{cv:energy-cubic-2d-upper-bound}
E^{\rm NLS}_{a_{n},b_{n}} \leq \inf_{\ell>0}\mathcal{E}_{a_{n},b_{n}}^{\rm NLS}\left[\ell^{\frac{1}{2}} Q_{0}\big(\ell^{\frac{1}{2}} \cdot\big)\right] & = \inf_{\ell>0} \left[-\ell \frac{a_{n} - a_{*}}{2} \|Q_{0}\|_{L^{4}}^{4} + \ell^{2} \frac{b_{n}}{6} \|Q_{0}\|_{L^{6}}^{6}\right] \nonumber \\
& = -\frac{3\|Q_{0}\|_{L^{4}}^{8}(a_{n} - a_{*})^{2}}{8\|Q_{0}\|_{L^{6}}^{6} b_{n}}.
\end{align}
This yields the energy upper bound in \eqref{cv:energy-cubic-2d}. We prove the matching energy lower bound in \eqref{cv:energy-cubic-2d} by showing the convergence of minimizers in \eqref{cv:minimizer-cubic-2d}. For this purpose, we set
$$
w_{n}(x) := \ell_{n}^{\frac{1}{2}} u_{n}\left(\ell_{n}^{\frac{1}{2}} x\right)
$$
where $\ell_{n}$ is given in \eqref{cv:minimizer-cubic-2d}. We rewrite the energy functional as follows
\begin{align}
E^{\rm NLS}_{a_{n}, b_{n}} = \mathcal{E}^{\rm NLS}_{a_{n}, b_{n}} [u_{n}] & = \ell_{n}^{-1}\left(\|\nabla w_{n}\|_{L^{2}}^{2} - \frac{a_{n}}{2} \|w_{n}\|_{L^{4}}^{4} \right) + \ell_{n}^{-2}\frac{b_{n}}{6}\|w_{n}\|_{L^{6}}^{6} \label{cv:energy-cubic-2d-lower-bound-1} \\
& \geq -\ell_{n}^{-1}\frac{a_{n}-a_{*}}{2}\|w_{n}\|_{L^{4}}^{4} + \ell_{n}^{-2}\frac{b_{n}}{6}\|w_{n}\|_{L^{6}}^{6}, \label{cv:energy-cubic-2d-lower-bound-2}
\end{align}
where we have used \eqref{ineq:gn}. It follows from \eqref{cv:energy-cubic-2d-upper-bound} and \eqref{cv:energy-cubic-2d-lower-bound-2} that
\begin{equation}\label{cv:energy-cubic-2d-boundedness}
-1 \geq -2\|Q_{0}\|_{L^{4}}^{-4} \|w_{n}\|_{L^{4}}^{4} + \|Q_{0}\|_{L^{6}}^{-6} \|w_{n}\|_{L^{6}}^{6}.
\end{equation}
Since $\|w_{n}\|_{L^{2}}^{2}=1$ and $\|w_{n}\|_{L^{4}}^{4} \leq \|w_{n}\|_{L^{6}}^{3}$, by H\"older inequality, we deduce from \eqref{cv:energy-cubic-2d-boundedness} that $\|w_{n}\|_{L^{4}}$ as well as $\|w_{n}\|_{L^{6}}$ are bounded uniformly from above and below. We use this to show the uniform boundedness of $\|\nabla w_{n}\|_{L^{2}}$.

Looking back at \eqref{cv:energy-cubic-2d-lower-bound-1}, by multiplying both sides by $\ell_{n}$, neglecting the nonnegative three-body interaction, using again \eqref{cv:energy-cubic-2d-upper-bound}, and taking the limit $n\to\infty$, we obtain
\begin{equation}\label{cv:minimizer-cubic-2d-profile}
0 \geq \lim_{n\to+\infty}\|\nabla w_{n}\|_{L^{2}}^{2} - \frac{a_{n}}{2} \|w_{n}\|_{L^{4}}^{4} = \lim_{n\to+\infty} \|\nabla w_{n}\|_{L^{2}}^{2} - \frac{a_{*}}{2} \|w_{n}\|_{L^{4}}^{4} \geq 0.
\end{equation}
Here we have used \eqref{ineq:gn} in the last inequality. Therefore, the equality in \eqref{cv:minimizer-cubic-2d-profile} must occur and this implies in particular the uniform boundedness of $\{w_{n}\}$ in $H^1(\mathbb{R}^{2})$. By Sobolev embedding, $\{w_{n}\}$ is also bounded uniformly in $L^{p}(\mathbb{ R}^{2})$, for every $2\leq p<\infty$. Note that $\{w_{n}\}$ is radially symmetric decreasing since $\{u_{n}\}$ is. By the same argument in the proof of \Cref{thm:existence}, we have that (up to a translation and extracting a subsequence) $w_{n} \to w_{0}$ weakly in $H^{1}(\mathbb{ R}^{2})$, strongly in $L^{p}(\mathbb R^{2})$ for every $2 <p<\infty$, and pointwise almost everywhere in $\mathbb R^{2}$. We prove that this is really strong convergence in $H^{1}(\mathbb R^{2})$. We first claim that $w_{0} \not\equiv 0$ since otherwise we get a contradiction, by \eqref{cv:energy-cubic-2d-boundedness}. Now, looking back at \eqref{cv:minimizer-cubic-2d-profile}, we use Fatou's lemma and \eqref{ineq:gn} to obtain
$$
0 = \lim_{n\to+\infty} \|\nabla w_{n}\|_{L^{2}}^{2} - \frac{a_{*}}{2} \|w_{n}\|_{L^{4}}^{4} \geq \|\nabla w_{0}\|_{L^{2}}^{2} - \frac{a_{*}}{2} \|w_{0}\|_{L^{4}}^{4} \geq \left(1-\|w_{0}\|_{L^{2}}^{2}\right) \|\nabla w_{0}\|_{L^{2}}^{2} \geq 0
$$
where we have used the fact that $0 < \|w_{0}\|_{L^{2}}^{2} \leq 1$ (by Fatou's lemma and the weak convergence $w_{n} \wto w_{0}$ in $L^{2}(\mathbb R^{2})$). Therefore, the equality must occur in the above and we must have that $\|w_{0}\|_{L^{2}}^{2} = 1$ and that $\|\nabla w_{n}\|_{L^{2}}^{2} \to \|\nabla w_{0}\|_{L^{2}}^{2}$. By Br\'ezis--Lieb lemma, $w_{n} \to w_{0}$ strongly in $H^{1}(\mathbb R^{2})$. Furthermore, it also yields that $w_{0}$ is an optimizer for \eqref{ineq:gn} which is unique (up to translation and dilation). Up to a translation, $w_{0}(x) = \sqrt{t}Q_0(\sqrt{t}x)$ where $t>0$ and $Q_{0}$ was chosen in \eqref{cv:energy-cubic-2d-upper-bound}. To complete the proof, we prove that $t = 1$ and hence $w_{0} \equiv Q_0$. Indeed, by taking the limit $n\to\infty$ in \eqref{cv:energy-cubic-2d-boundedness}, we obtain
\[
-1 \geq -2\|Q_{0}\|_{L^{4}}^{-4} \|w_{0}\|_{L^{4}}^{4} + \|Q_{0}\|_{L^{6}}^{-6} \|w_{0}\|_{L^{6}}^{6} = -2t +t^{2} \geq -1. 
\]
Here we have used the Cauchy--Schwarz inequality in the last inequality. Obviously, the equality in the above occurs at $t=1$. Finally, the convergence \eqref{cv:minimizer-cubic-2d} holds for the whole sequence since the limiting profile $Q_{0}$ is unique. The proof is completed.


\begin{thebibliography}{10}

\bibitem{AkhDasVag-99}
{\sc N.~Akhmediev, M.~P. Das, and A.~Vagov}, {\em {Bose-Einstein condensation
  of atoms with attractive interaction}}, International Journal of Modern
  Physics B, 13 (1999), pp.~625--631.

\bibitem{BisBla-15}
{\sc R.~Bisset and P.~Blakie}, {\em {Crystallization of a dilute atomic dipolar
  condensate}}, Physical Review A, 92 (2015), p.~061603.

\bibitem{Blakie-16}
{\sc P.~B. Blakie}, {\em {Properties of a dipolar condensate with three-body
  interactions}}, Physical Review A, 93 (2016), p.~033644.

\bibitem{BloDalWil-08}
{\sc I.~Bloch, J.~Dalibard, and W.~Zwerger}, {\em {Many-body physics with
  ultracold gases}}, Reviews of modern physics, 80 (2008), pp.~885--964.

\bibitem{BraSacHul-97}
{\sc C.~C. Bradley, C.~Sackett, and R.~Hulet}, {\em {Bose-Einstein condensation
  of lithium: Observation of limited condensate number}}, Physical Review
  Letters, 78 (1997), p.~985.

\bibitem{BraSacTolHul-95}
{\sc C.~C. Bradley, C.~Sackett, J.~Tollett, and R.~G. Hulet}, {\em {Evidence of
  Bose-Einstein condensation in an atomic gas with attractive interactions}},
  Physical review letters, 75 (1995), p.~1687.

\bibitem{BreLie-83}
{\sc H.~Brezis and E.~H. Lieb}, {\em {A Relation Between Pointwise Convergence
  of Functions and Convergence of Functionals}}, Proceedings of the American
  Mathematical Society, 88 (1983), pp.~486--490.

\bibitem{ChePav-11}
{\sc T.~Chen and N.~Pavlovi{\'c}}, {\em {The quintic NLS as the mean field
  limit of a Boson gas with three-body interactions}}, Journal of Functional
  Analysis, 260 (2011), pp.~959--997.

\bibitem{Chen-12}
{\sc X.~Chen}, {\em {Second order corrections to mean field evolution for
  weakly interacting bosons in the case of three-body interactions}}, Archive
  for Rational Mechanics and Analysis, 203 (2012), pp.~455--497.

\bibitem{CheHol-13}
{\sc X.~Chen and J.~Holmer}, {\em {Focusing Quantum Many-body Dynamics: The
  Rigorous Derivation of the 1D Focusing Cubic Nonlinear {S}chr{\"o}dinger
  Equation}}, Archive for Rational Mechanics and Analysis, 221 (2016),
  pp.~631--676.

\bibitem{CheHol-17}
\leavevmode\vrule height 2pt depth -1.6pt width 23pt, {\em {The rigorous
  derivation of the 2D cubic focusing NLS from quantum many-body evolution}},
  International Mathematics Research Notices, 2017 (2017), pp.~4173--4216.

\bibitem{CheHol-19}
\leavevmode\vrule height 2pt depth -1.6pt width 23pt, {\em {The derivation of
  the $\mathbb{T}^3$ energy-critical NLS from quantum many-body dynamics}},
  Inventiones Mathematicae, 217 (2019), pp.~433--547.

\bibitem{DalGioPitStr-99}
{\sc F.~Dalfovo, S.~Giorgini, L.~P. {P}itaevskii, and S.~Stringari}, {\em
  {Theory of {B}ose--{E}instein condensation in trapped gases}}, Reviews of
  Modern Physics, 71 (1999), pp.~463--512.

\bibitem{DinNguRou-24}
{\sc V.~D. Dinh, D.-T. Nguyen, and N.~Rougerie}, {\em {Blowup of
  two-dimensional attractive Bose--Einstein condensates at the critical
  rotational speed}}, Annales de l'Institut Henri Poincar{\'e} C, 41 (2024),
  pp.~1055--1081.

\bibitem{EsrGreZhoLin-96}
{\sc B.~Esry, C.~H. Greene, Y.~Zhou, and C.~Lin}, {\em Role of the scattering
  length in three-boson dynamics and bose-einstein condensation}, Journal of
  Physics B: Atomic, Molecular and Optical Physics, 29 (1996), p.~L51.

\bibitem{GamFreTom-99}
{\sc A.~Gammal, T.~Frederico, and L.~Tomio}, {\em {Trapped Bose-Einstein
  condensed gas with two and three-atom interactions}}, in Proceedings of the
  International Workshop, {C. Bertulani, LF. Canto and M. Hussein}, ed., World
  Scientific, 1999.

\bibitem{GamFreTomCho-00}
{\sc A.~Gammal, T.~Frederico, L.~Tomio, and P.~Chomaz}, {\em {Atomic
  Bose-Einstein condensation with three-body interactions and collective
  excitations}}, Journal of Physics B: Atomic, Molecular and Optical Physics,
  33 (2000), p.~4053.

\bibitem{GuoSei-14}
{\sc Y.~Guo and R.~Seiringer}, {\em {On the Mass Concentration for
  {B}ose--{E}instein Condensates with Attractive Interactions}}, Letters in
  Mathematical Physics, 104 (2014), pp.~141--156.

\bibitem{JosRic-97}
{\sc C.~Josserand and S.~Rica}, {\em {Coalescence and droplets in the
  subcritical nonlinear Schr{\"o}dinger equation}}, Physical Review Letters, 78
  (1997), p.~1215.

\bibitem{KinWenWei-04}
{\sc T.~Kinoshita, T.~Wenger, and D.~S. Weiss}, {\em {Observation of a
  One-Dimensional {T}onks-{G}irardeau Gas}}, Science, 305 (2004),
  pp.~1125--1128.

\bibitem{LewNamRou-17-proc}
{\sc M.~Lewin, P.~T. Nam, and N.~Rougerie}, {\em Blow-up profile of rotating 2d
  focusing {B}ose gases}, in Workshop on Macroscopic Limits of Quantum Systems,
  Springer, 2017, pp.~145--170.

\bibitem{LewNamRou-17}
\leavevmode\vrule height 2pt depth -1.6pt width 23pt, {\em A note on 2d
  focusing many-boson systems}, Proceedings of the American Mathematical
  Society, 145 (2017), pp.~2441--2454.

\bibitem{LiYao-21}
{\sc Y.~Li and F.~Yao}, {\em {Derivation of the nonlinear Schr{\"o}dinger
  equation with a general nonlinearity and Gross--Pitaevskii hierarchy in one
  and two dimensions}}, Journal of Mathematical Physics, 62 (2021), p.~021505.

\bibitem{LieLos-01}
{\sc E.~H. Lieb and M.~Loss}, {\em {Analysis}}, vol.~14 of {Graduate Studies in
  Mathematics}, American Mathematical Society, Providence, RI, 2nd~ed., 2001.

\bibitem{LieSeiSolYng-05}
{\sc E.~H. Lieb, R.~Seiringer, J.~P. Solovej, and J.~Yngvason}, {\em {The
  mathematics of the {B}ose gas and its condensation}}, {Oberwolfach
  {S}eminars}, Birkh{\"a}user, 2005.

\bibitem{LieSeiYng-00}
{\sc E.~H. Lieb, R.~Seiringer, and J.~Yngvason}, {\em {Bosons in a trap: A
  rigorous derivation of the {G}ross--{P}itaevskii energy functional}},
  Physical Review A, 61 (2000), p.~043602.

\bibitem{LieSeiYng-01}
{\sc E.~H. Lieb, R.~Seiringer, and J.~Yngvason}, {\em {A Rigorous Derivation of
  the Gross--Pitaevskii Energy Functional for a Two-dimensional Bose Gas}},
  Communications in Mathematical Physics, 224 (2001), pp.~17--31.

\bibitem{NamRicTri-22b}
{\sc P.~T. Nam, J.~Ricaud, and A.~Triay}, {\em {Dilute Bose gas with three-body
  interaction: Recent results and open questions}}, Journal of Mathematical
  Physics, 63 (2022), p.~061103.

\bibitem{NamRicTri-22a}
\leavevmode\vrule height 2pt depth -1.6pt width 23pt, {\em {Ground state energy
  of the low density Bose gas with three-body interactions}}, Journal of
  Mathematical Physics, 63 (2022), p.~071903.

\bibitem{NamRicTri-23}
\leavevmode\vrule height 2pt depth -1.6pt width 23pt, {\em {The condensation of
  a trapped dilute Bose gas with three-body interactions}}, Probability and
  Mathematical Physics, 4 (2023), pp.~91--149.

\bibitem{NamRouSei-16}
{\sc P.~T. Nam, N.~Rougerie, and R.~Seiringer}, {\em {Ground states of large
  Bose systems: The Gross--Pitaevskii limit revisited}}, Analysis and PDEs, 9
  (2016), pp.~459--485.

\bibitem{NamSal-20}
{\sc P.~T. Nam and R.~Salzmann}, {\em {Derivation of 3D energy-critical
  nonlinear Schr{\"o}dinger equation and Bogoliubov excitations for Bose
  gases}}, Communications in Mathematical Physics, 375 (2020), pp.~495--571.

\bibitem{NguRic-24}
{\sc D.-T. Nguyen and J.~Ricaud}, {\em On one-dimensional bose gases with
  two-body and (critical) attractive three-body interactions}, SIAM Journal on
  Mathematical Analysis, 56 (2024), pp.~3203--3251.

\bibitem{NguRic-25}
\leavevmode\vrule height 2pt depth -1.6pt width 23pt, {\em {Stabilization
  against collapse of 2D attractive Bose--Einstein condensates with repulsive,
  three-body interactions}}, Letters in Mathematical Physics, 115 (2025),
  pp.~1--46.

\bibitem{NguRou-22}
{\sc D.-T. Nguyen and N.~Rougerie}, {\em {Thomas--Fermi profile of a fast
  rotating Bose--Einstein condensate}}, Pure and Applied Analysis, 4 (2022),
  pp.~535--569.

\bibitem{ParWidMurManFolCirGorShlTheHanBlo-04}
{\sc B.~Paredes, A.~Widera, V.~Murg, O.~Mandel, S.~F{\"o}lling, I.~Cirac, G.~V.
  Shlyapnikov, T.~W. H{\"a}nsch, and I.~Bloch}, {\em {Tonks--Girardeau gas of
  ultracold atoms in an optical lattice}}, Nature, 429 (2004), pp.~277--281.

\bibitem{Petrov-14}
{\sc D.~Petrov}, {\em {Three-body interacting bosons in free space}}, Physical
  Review Letters, 112 (2014), p.~103201.

\bibitem{PetHolShl-00}
{\sc D.~Petrov, M.~Holzmann, and G.~Shlyapnikov}, {\em {Bose-Einstein
  condensation in quasi-2D trapped gases}}, Physical Review Letters, 84 (2000),
  p.~2551.

\bibitem{SacStoHul-98}
{\sc C.~Sackett, H.~Stoof, and R.~Hulet}, {\em {Growth and collapse of a
  Bose--Einstein condensate with attractive interactions}}, Physical Review
  Letters, 80 (1998), p.~2031.

\bibitem{SchVin-00}
{\sc P.~Schuck and X.~Vinas}, {\em {Thomas-Fermi approximation for
  Bose-Einstein condensates in traps}}, Physical Review A, 61 (2000),
  p.~043603.

\bibitem{SulSulPat-84}
{\sc P.~Sulem, C.~Sulem, and A.~Patera}, {\em {Numerical simulation of singular
  solutions to the two-dimensional cubic Schr{\"o}dinger equation}},
  Communications on pure and applied mathematics, 37 (1984), pp.~755--778.

\bibitem{Weinstein-83}
{\sc M.~I. Weinstein}, {\em {Nonlinear {S}chr{\"o}dinger equations and sharp
  interpolation estimates}}, Communications in Mathematical Physics, 87 (1983),
  pp.~567--576.

\bibitem{Xie-15}
{\sc Z.~Xie}, {\em {Derivation of a nonlinear Schr{\"o}dinger equation with a
  general power-type nonlinearity in $ d= 1, 2$}}, Differential and Integral
  Equations, 28 (2015), pp.~455--504.

\bibitem{Yuan-15}
{\sc J.~Yuan}, {\em {Derivation of the Quintic NLS from many-body quantum
  dynamics in $\mathbb T^2$}}, Communications on Pure \& Applied Analysis, 14
  (2015), p.~1941.

\end{thebibliography}

\end{document}